\documentclass[]{elsart}

\usepackage{amssymb}
\usepackage{graphicx}
\usepackage{dcolumn}
\usepackage{bm}
\usepackage{subfigure}
\begin{document}

\begin{frontmatter}



\title{Dynamics and Topological Aspects of a Reconstructed Two-Dimensional Foam
Time Series Using Potts Model on a Pinned Lattice}

\author[ND]{Igor F. Vasconcelos},
\author[UR1]{Isabelle Cantat}, and
\author[IUB]{James A. Glazier}

\address[ND]{Department of Physics, University of Notre Dame, 225 Nieuwland Science Hall,
46556-5670, Notre Dame, IN, USA}
\address[UR1]{GMCM, U.M.R. C.N.R.S. 6626, Campus de Beaulieu,  B\^atiment 11A, CS
74205 263, av. du G\'en\'eral Leclerc, 35042 Rennes Cedex, France}
\address[IUB]{Department of Physics, Indiana University, Swain Hall West 159, 727
East Third Street, 47405-7105, Bloomington, IN, USA}

\begin{abstract}
We discuss a method to reconstruct an approximate two-dimensional
foam structure from an incomplete image using the extended Potts
model on a pinned lattice. The initial information consists of the
positions of the vertices only. We locate the centers of the
bubbles using the Euclidean distance-map construction and assign
at each vertex position a continuous pinning field with a
potential falling off as $1/r$. We nucleate a bubble at each
center using the extended Potts model and let the structure evolve
under the constraint of scaled target areas until the bubbles
contact each other. The target area constraint and pinning centers
prevent further coarsening. We then turn the area constraint off
and let the edges relax to a minimum energy configuration. The
result is a reconstructed structure very close to the simulation.
We repeated this procedure for various stages of the coarsening of
the same simulated foam and investigated the simulation and
reconstruction dynamics, topology and area distribution, finding
that they agreed to good accuracy.
\end{abstract}

\end{frontmatter}

\section{Introduction}

This paper addresses the reconstruction of domain boundaries from
experimental images of foams, cellular images which consist of
sets of compact domains separated from each other by sharp
boundaries which meet at vertices. The data to be reconstructed
contain only partial information, in particular the positions of
vertices (two dimensions) or edges (three dimensions) in the complete image. This task,
which is a special case of the problem of image segmentation,
recurs in many fields, {\it e.g.} medical imaging and
cartography~\cite{Munoz03,Pham00,Suri02}. An ISI Web of Science
search reveals nearly 2000 references on image segmentation. Even
when the image data is complete, a typical voxel image requires
reconstruction to obtain information about domain shapes, volumes
and adjacencies. In real images the presence of noise further
complicates reconstruction~\cite{Zindzindohoue91}.

The general reconstruction problem is underdetermined. Unless we
supply further constraints, an arbitrary number of different
domain patterns with different numbers of domains and domain area
distributions can have the same set of vertices or edges. 
As a result, almost all reconstruction of
three-dimensional foams has been manual, which is inconvenient and
time consuming and greatly limits the numbers of bubbles which we
can examine~\cite{prause00,monnereau98a,monnereau98b,monnereau99,monnereau00,monnereau01,fetterman00}.

However,
in the case of many domain patterns of real interest, {\it e.g.}
metallic polycrystals, foams and biological cells, the patterns
are well behaved in a manner which is clear to the eye, though
difficult to quantify completely. For example, in two dimensions,
boundaries meet at three-fold vertices at approximately
120$^\circ$ angles and are relatively smooth and constant in
curvature. Many of these regularities, known as Plateau
rules~\cite{plateau}, result from the physical processes
generating the domain boundaries: the presence of a surface
tension or surface energy causes boundaries to be approximately
minimal surfaces. 

In these cases we can proceed quite far in reconstructing
the complete two-dimensional pattern from the vertex positions and we present
an algorithm that accomplishes this reconstruction with accuracy
good enough for many practical purposes. The algorithm is
attractive in that it uses a physical simulation of the
development of minimal surfaces which in some sense duplicates the
original formation of the pattern. One great advantage is that it
works directly with real pixel-wise input data. We illustrate
several pathological cases where the algorithm fails and discuss
ways to fix these errors.

\section{Motivation}

Why do we need an algorithm that can reconstruct minimal surface partitions from
vertex positions? Whole subfields of physics and mathematics have developed to study
the structure of complex minimal surfaces (for a review see the web site of Kenneth Brakke and
his program Surface Evolver at http://www.susqu.edu/facstaff/b/brakke/).

The division of space into subdomains with minimum partition area has attracted
attention of scientists in fields ranging from Geography and Physics to Mathematics and Computer
Science~\cite{Morgan02,Morgan96}. The more restricted question of which
geometric structure, replicated infinitely and filling space without leaving gaps, has
minimum surface area, is known as the Kelvin Problem after Lord Kelvin who first proposed a
solution in 1887~\cite{kelvin87}.

Dry liquid foams ({\it i.e.} foams in which the liquid fraction is below about 1\%) are a good probe of
the various proposed solutions to the Kelvin problem. The idealized model of a foam retains only the
surface energy (proportional to the surface area) of the bubbles. Where bubbles
contact each other, they form a cell face of negligible thickness. The bubble wall membranes bend, forming
surfaces of constant mean curvature. The faces contact each other at edges known as {\it Plateau borders} where water
accumulates and these edges in turn meet at {\it Plateau vertices} at the tetrahedral angle.

Foams also exhibit a complex dynamics in which gas passes from
bubbles of higher pressure to lower pressure
bubbles~\cite{weaire84,atkinson88}. Since smaller bubbles tend to
have higher pressures because of geometric constraints, bubbles
gradually disappear and the pattern
coarsens~\cite{weaire84,atkinson88}. The same general phenomena
with additional complications occur in metallic polycrystals,
sintered ceramics, ferromagnets, micelles, ferrofluids and
biological tissues and
organs~\cite{weaire84,reiter92,thompson42,dormer80}. Domain growth
is a common feature of most of these patterns, though the time
scales differ depending on the diffusion mechanisms controlling
boundary motion. If the initial conditions are reasonably
homogeneous, most of these materials develop geometrically similar
structures~\cite{fradkov85}. Thus, understanding the kinetics and
geometry of a coarsening foam can provide information on a broad
class of important materials.

\begin{figure}
\centering
\includegraphics[width=0.6\linewidth]{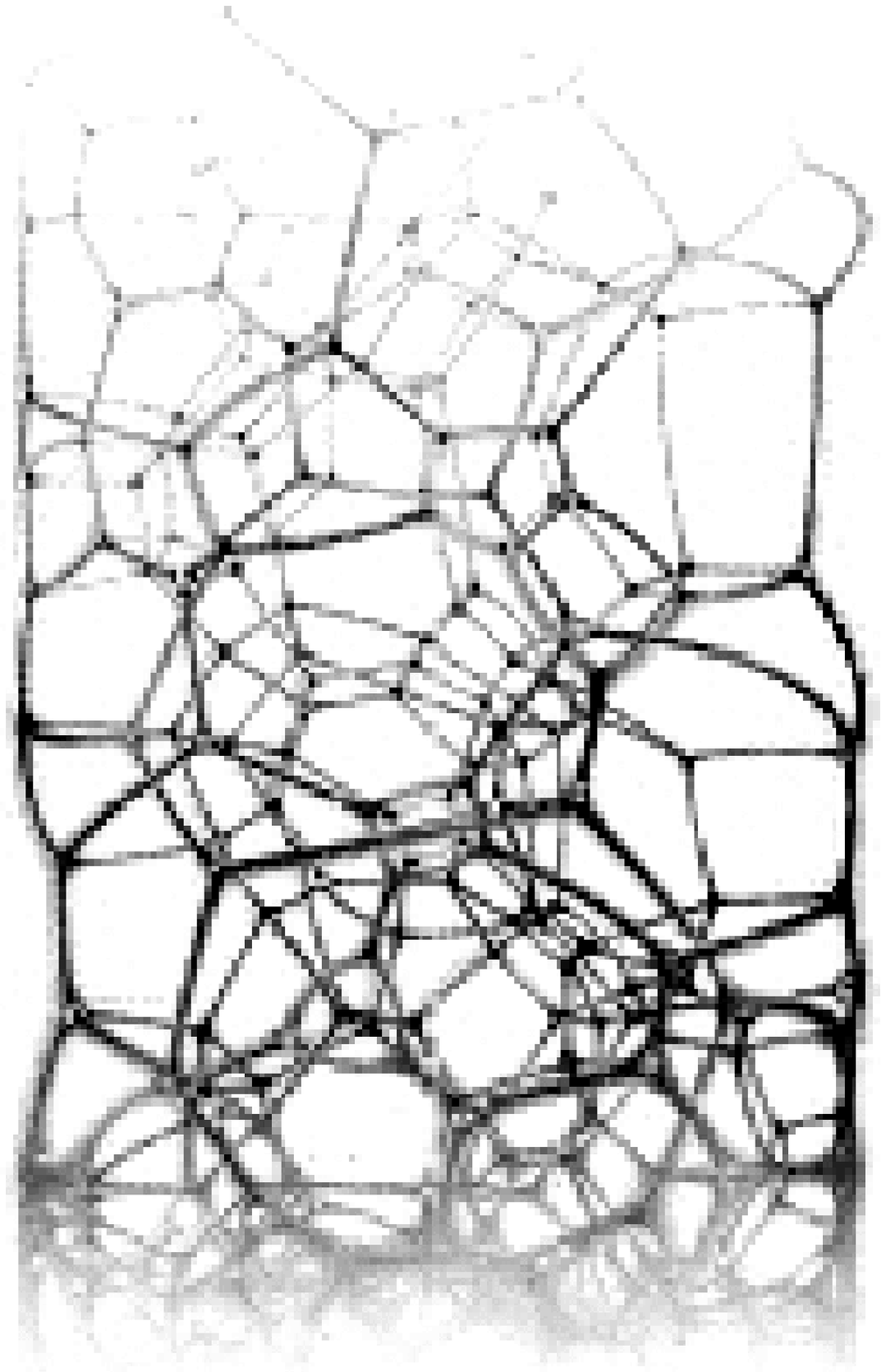}
\caption{MRI image of a foam from~\cite{prause00}.}
\label{MRI}
\end{figure}

Unfortunately, determining the three-dimensional structure of a
foam is surprisingly difficult. Foams scatter light extremely
effectively, so seeing deep inside a liquid foam is
difficult~\cite{durian91a,durian91b}. Optical tomography is only
possible for very dry foams, in which the Plateau borders but not
the faces are
visible~\cite{monnereau98a,monnereau98b,monnereau99,monnereau00,monnereau01,fetterman00}.
Magnetic resonance imaging (MRI) can also reveal a foam's
three-dimensional structure~\cite{gonatas95}. MRI measures the
nuclear magnetic resonance (NMR) signal from protons in the
hydrogen in the water present in the sample. However, since foams
contain very little water, which mostly concentrates in the
Plateau borders, the MRI signal is very weak and the best result
we obtain is an image of the Plateau borders, which looks like a
skeleton of the foam as in fig.~\ref{MRI}. Recent synchrotron CT
images provide higher quality images~\cite{graner02}, but seeing
the faces between bubbles remains effectively impossible.

The raw image is missing crucial information, {\it e.g.} the
bubble volumes, because of the absence of the bubble membranes.
The same problem can occur in confocal microscopy of cells, CT,
MRI and ultrasound imaging of tissues and X-ray and neutron
imaging of solids. An added difficulty is that the signal-to-noise
ratio of such images is often close to one.

Reconstruction of foam structure from partial information is,
then, essential to the study of cellular patterns. The concept
underlying our reconstruction is energy minimization, in the sense
that reconstructing the pattern is equivalent to finding the
minimum energy partition consistent with the partial information
available about the structure. Since the pattern itself results
from constrained energy minimization, the use of an energy
minimization method in reconstruction seems natural and appealing.
In three dimensions, this partial information consists of a pixel
image of the edges of bubbles (Plateau borders) and the
reconstruction consists of finding the surfaces that connect the
edges with minimum energy. In two dimensions, on the other hand,
the corresponding partial data are the positions of the vertices
of the domains and the reconstruction connects the vertices to
obtain a minimum energy configuration. Although the results we
present here concern two-dimensional reconstruction only, the
method extends to three dimensions without major conceptual
modifications.

We now summarize our method to reconstruct the two-dimensional
structure starting solely from the positions of the vertices and
our knowledge of minimal surfaces. Subsequent sections provide
details and results. The method consists of three main independent
steps:
\begin{itemize}
\item Finding the centers of bubbles in the pattern to be
reconstructed. \item Reconstructing the pattern by nucleating a
bubble from each center using the extended large-$Q$ Potts Model.
\item Growing the nucleated bubbles until evolution stops.
\end{itemize}
Minimum distance maps provide the positions of the centers and the
initial target areas (Each bubble receives a target area which we
adjust during the simulations to achieve the minimum surface
energy for each bubble)~\cite{Glantz97}. A continuous pinning
field traps the moving bubble edges.

We employ this method to reconstruct several patterns from a foam
evolution time series obtained from a coarsening simulation. In
order to test the validity of our results, we perform a detailed
analysis of the dynamics and spatial structure of the
reconstructed foam and compare it with the simulation and
theoretical and experimental results found in the literature.

\section{The Potts Model}

The Potts Model~\cite{wu82} is a generalization of the Ising model
to more than two spin components. Although initially proposed to
study critical phenomena in statistical physics, the Potts Model
finds a wide variety of applications, including the simulation of
metallic grain
growth~\cite{srolovitz83,srolovitz84a,srolovitz84b}, soap
foams~\cite{wejchert86,glazier90,jiang96}, magnetic
froths~\cite{weaire91} and biological cells~\cite{graner92}.

We map the foam structure onto a rectangular lattice containing
$N_{x}*N_{y}$ sites. Each lattice site contains an integer
$\sigma$ (a spin) which corresponds to a particular bubble (number
of bubbles = number of spins). The boundary between two bubbles is
the set of links between lattice sites associated to the spins of
those bubbles. The boundary energy associates a positive energy
(equivalent to a surface tension) with boundary links and zero
energy for links within bubbles.

Coarsening
has two time scales. Surface tension and viscous dissipation
determine the time, $\tau_1$, which describes the relaxation of an
edge towards equilibrium without volume (or area in two dimensions)
variation. After a perturbation, the pattern returns to obeying
the Plateau rules after a delay of $\tau_1$. The second time
scale, $\tau_2$, depends mainly on gas permeability and foam
polydispersity and controls the coarsening rate. In liquid foams
$\tau_1$ is typically fractions of a second while $\tau_2$ is
minutes or hours. In metals the two times are usually comparable,
as they are in the simple Potts Model, so the foam only obeys
Plateau's rules approximately. Wejchert {\it et al.}~\cite{wejchert86} introduced an area
constraint energy term to simulate equilibration in a foam, where
boundary equilibration is much faster than diffusion. If we constrain the area for a
fixed number of Monte Carlo steps ({\it mcs} as defined below) per volume diffusion
step we can choose any ratio we like between the time scales. 

The Hamiltonian including surface energy and the area constraint has the general form:
\begin{equation}
H= J \sum_{i,j (sites)}(1-\delta_{\sigma_i
\sigma_j})+\lambda\sum_{k (bubbles)}(S_k - S_k^t)^2 \textrm{.}
\label{hamilgeral}
\end{equation}
The first term in eq.~\ref{hamilgeral} accounts for the surface
energy. The sum in $i$ runs over all the sites in the lattice and
the sum in $j$ runs over sites neighboring $i$. The parameter $J$
sets the energy/unit surface area. The second term is the area
constraint. The sum in $k$ runs over all the bubbles in the
pattern. $\lambda$ is a parameter specifying the strength of the
area constraint, $S_k$ the current area of the $k^{th}$ bubble,
and $S_k^t$ the target area of the same bubble. Because of the
surface energy, each bubble's area is usually smaller than its
target area.

Most Potts simulation use a modified Metropolis Monte Carlo
dynamics~\cite{metropolis53} in which we select a boundary lattice
site at random and randomly propose to change its spin to the
value of one of the neighboring spins (Kawasaki
dynamics~\cite{kawasaki65} and Glauber dynamics~\cite{glauber63}
are also possible). If the resulting change of energy is less than
or equal to zero, we accept the new spin configuration. However,
if the change in energy is positive, we accept the new
configuration with Boltzmann probability:
\begin{equation}
P=\exp \left( -\frac{\Delta E}{kT} \right) \textrm{,}
\end{equation}
where $T$ is the temperature assigned to allows for thermal fluctuations to
overcome local energy minima. We
define a Monte Carlo step as a sequence of $N_{x}*N_{y}$ random
site selections.

\section{Generation of the input data}

We may use any experimental picture of a two-dimensional foam as input data, once
we know the vertex positions. To compare the statistical
properties of the reconstructed foam to those of the original
foam, we need a controlled series of well characterized foam
structures. To develop and test our reconstruction procedures, we
generate two-dimensional foam structures at different stages of a
coarsening evolution using the Large-Q Potts
Model~\cite{glazier90}.

The Hamiltonian contains the surface term only:
\begin{equation}
H = J \sum_{i,j}(1-\delta_{\sigma_i \sigma_j}) \textrm{.}
\end{equation}
The $j$ sum ranges up to third-nearest neighbors, or 20 sites
around the $i^{th}$ site. In our simulation $J=1$. In order to
minimize finite size and edge effects we use a $1024 \times 1024$
square lattice with periodic boundary conditions. We start the
simulations with $16384$ square $8 \times 8$ bubbles and finish
with just a few tens of bubbles.

We performed the main part of the numerical simulation at $T= 0$.
However, in order to overcome metastable traps, we periodically
increased the temperature for short periods during the simulation, alternating
40 {\it mcs} at temperature $T=0$ (relaxation
period) with 5 {\it mcs} at $T= 0.5$ (fluctuation period). This
method produces very realistic foam structures~\cite{glazier90}.
Thus the Potts Model basis of both original and reconstructed
structures does not bias our error estimates.

\section{Pinning Field}

Our inspiration for a continuous attractive pinning field to keep
vertices fixed comes from Zener
pins~\cite{srolovitz84c,kad97,soucail99,miyake01,krichevsky92,herrera97},
which simulate the microstructural evolution of materials in the
presence of a second phase dispersion of particles. The presence
of insoluble precipitates inhibits grain growth, greatly affecting
the mechanical properties of the material.

Srolovitz {\it et al.}~\cite{srolovitz84c} extended the Potts
Model to simulate grain growth in the presence of a second phase
particle dispersion. Their model incorporates particles by
assigning a particular spin to the lattice sites the particles
occupy. The kinetics is the regular Monte Carlo dynamics for the
Potts Model. Other groups~\cite{kad97,soucail99,miyake01} have
also used this procedure to obtain properties including grain
size, characteristic exponents, {\it etc.} for such pinned
materials. Krichevsky and Stavans~\cite{krichevsky92} and Herrera
{\it et al.}~\cite{herrera97} performed two-dimensional
experiments on foam evolution in the presence of pinning centers,
using a square array of pins and randomly distributed pins
respectively.

We construct a continuous attractive pinning field around a point
size Zener pin at each vertex in the source image, contributing a
Hamiltonian term:
\begin{equation}
H_p = -\gamma \sum_{i,j (sites)}(1-\delta_{\sigma_i \sigma_j})\textrm{F}(x_i,y_i)=
        -\gamma \sum_{i,j (sites)}(1-\delta_{\sigma_i \sigma_j})\times\bigg(\frac{1}{r_i}\bigg)
\textrm{,} \label{eqpin}
\end{equation}
where $\gamma$ is a constant that determines the pinning coupling
strength and $r_i$ is the distance from the site $i$ to the
closest pin. This term causes edges, and ultimately the
reconstructed vertices, to move towards and eventually trap on the
pins. The field F$(x,y)$ has its maximum value of $1$ at the
position of the pins and falls off as the inverse of the distance
($1/r$). It contributes a negative energy at each place where a
link between two different bubbles is present (see
eq.~\ref{eqpin}). Therefore, boundaries and vertices cost less
energy when located close to a pinning center, producing an
attractive pinning force on the vertices.

We choose the value of $\gamma$ to produce a circular region
around each center where the attraction of that center dominates,
{\it i.e.}, the center will definitely attract any edge or vertex
within this circle. In our simulations, we chose the circles'
radii to be 20 pixels, which gives a value for the field, at the
edge of the circles, of about 0.05. With $\gamma= 1000$, the
pinning energies are strong enough to ensure that almost no edges
or vertices will slip off a pin. In the region between two pins
closer than 40 lattice sites, the closest pin will attract the
vertex due to the stronger pinning field in that direction. This
situation becomes less significant when we impose the cutoffs
discussed below.

\begin{figure}[ht]
\centering
\includegraphics[width=\linewidth]{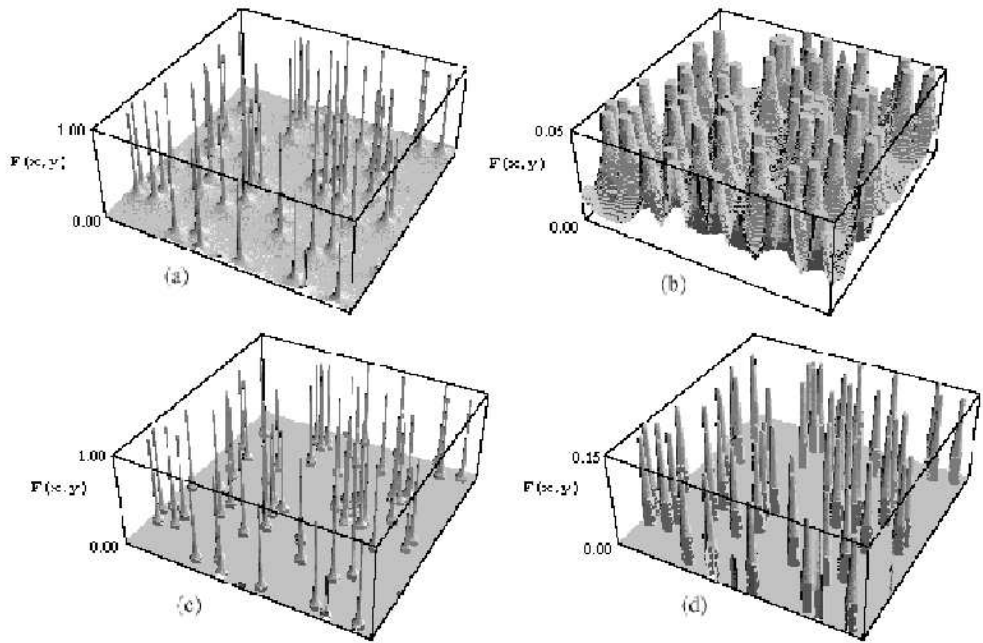}
\caption{(a) Pinning field for a particular foam configuration. The spikes sit at the positions of the vertices.
(b) The same pinning field scaled to show the circle around each pin as discussed in the text. (c) and (d) Pinning field for
the same foam configuration as in fig. 3 obtained after the cutoff discussed in the text. The vertical axis corresponds to
the field value $F(x,y)$ at the $(x,y)$ position in the lattice (horizontal plane). We obtained this foam configuration
from the coarsening simulation at 50000 {\it mcs}.}
\label{pinfield}
\end{figure}

Figure~\ref{pinfield}a shows the entire range of the pinning field
for a particular foam configuration while fig.~\ref{pinfield}b
shows the same field scaled to show the field within 20 pixels of
the pins. The pinning field within the circles (visible in the top
plane in fig.~\ref{pinfield}b) around each pin is much stronger
than other energy components. The great advantage of this method
compared to simple Zener pins is that in the final patterns almost
all pins lie in edges. If the nucleation algorithm has missed
bubbles, then the number of pins may be larger than the number of
vertices in the reconstructed pattern and a few pins may not
correspond to vertices. If the nucleation algorithm creates extra
bubbles then a few vertices may not correspond to pins.

A drawback of this implementation is that the lattice becomes very
stiff and, although the vertices pin appropriately, the edges are
not free to relax. To overcome this problem, we impose a cutoff on
the field, after a certain number of {\it mcs}. After the cutoff,
the field still falls off as $1/r$ but we limit its range to a
certain radius around each pin. This procedure keeps the vertices
pinned but frees most of the edge length from the pinning field,
allowing edges to straighten in order to minimize surface energy.
We apply successively shorter range cutoffs until the range of the
pinning field is about four lattice sites.
Figure~\ref{pinfield}c-d shows the field configuration after a
cutoff for $r=$ 20.

\section{Reconstruction}

\subsection{Centers of Bubbles}

\begin{figure}
\includegraphics[width=\linewidth]{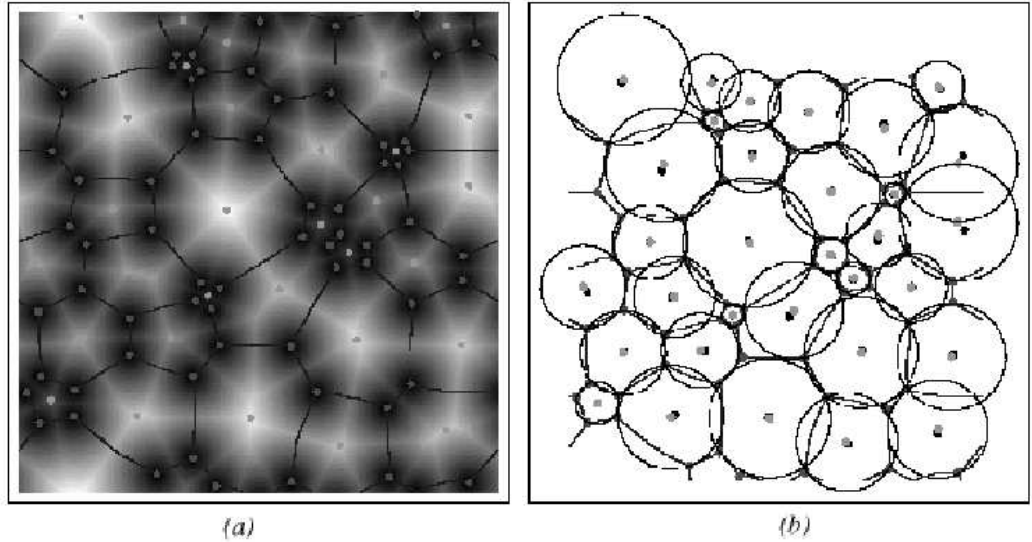}
\caption{(a) Minimum distance map superimposed on the original foam structure
and the centers obtained from the maxima. (b) Positions of calculated
centers (green) compared to the positions of the real ones (black). Pinning
centers are shown in red and the original structure in blue. The areas of te circles show our estimated target area for each bubble.
Both pictures correspond to the 5000 {\it mcs} coarsened foam structure.}
\label{maps}
\end{figure}

Our first step is to find the bubbles' centers. We follow Glantz
and Prause and use a Euclidean distance map to find the
centers~\cite{prause00,Glantz97,Glantz96,Glantz99,Blurock99,Timonen01,Hidajat02,Jang02,Kulountzakis92}.
We construct such a map by picking a lattice site, calculating its
distance from all the vertices and storing the minimum of these
distances for that site. Repeating this procedure for all lattice
sites produces the map~\cite{Kulountzakis92,chen94,fujiwara95}.
Figure~\ref{maps}a shows the map for a typical foam configuration.
We assume that the centers of the bubbles sit on the local maxima
in the map. Figure~\ref{maps}b shows the positions of the
calculated centers (green) compared to the real ones (black)
obtained from center of mass calculations. The agreement is fairly
good. In  three dimensions, this process is essentially the same, with the sole
change that the distances are calculated from the sites to the
edges.

\begin{figure}
\centering
\includegraphics[width=\linewidth]{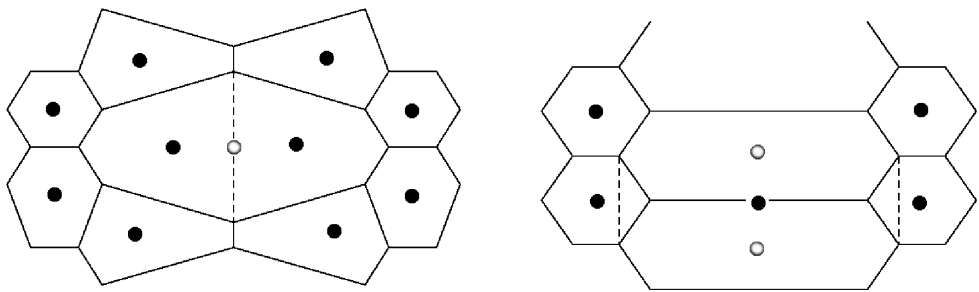}
\caption{Two typical errors occuring during bubble center determination. Bullets
show the bubble centers obtained with our modified
Euclidean-distance algorithm and circles show the actual centers. The solid
lines show the boundaries of the simulated input foam and the
dotted lines the reconstructed structure.}
\label{patho}
\end{figure}

This method, however, has some pathologies. Multiple maxima may be
present within a bubble. We treat this problem by combining any
two maxima that are closer to each other than to the closest
vertex. The solution has an intrinsic flaw: when a bubble is long
and thin, its centers may be closer to neighboring bubble centers
than to the closest vertex, causing the disappearance of the
bubble. Alternatively, the centers in a stretched bubble may be
further from each other than from the nearest edge, causing the
center map to create an extra bubble. Figure~\ref{patho} illustrates these two situations. 

Saddle points
are another problem. The center point of an edge is a saddle point
in the Euclidean map and if the edge is long enough, the saddle
point may look like a maximum because of errors resulting from
lattice discretization. An effort to exclude saddle points when
searching for maxima may result in missing centers for small
bubbles. If the initial bubble size distribution is narrow
(bubbles are roughly the same size) and the domains are fairly
isotropic, then no errors result. We are currently working to
quantify the consequences and thresholds of these errors. Overall,
we lose a few very small or elongated bubbles. During foam
coarsening, these bubbles have a short lifetime (they disappear
rapidly), so the error in most cases is quite small and has the
effect of reconstructing a foam pattern that seems slightly later
in time than the source pattern.

These pathologies result from the distance algorithm, {\it i.e.},
the first part of our method. Our main reconstruction procedure,
however, is independent of the method we use to find the centers
of bubbles which we can improve or replace, {\it e.g.} the method proposed by
Kammerer and Glantz~\cite{kammerer03}.

\subsection{Nucleation}

Once we have chosen a set of bubble centers we nucleate a bubble
with unique spin and area of one pixel at each center and set the
remaining sites to spin 0. The spin 0 domain behaves like a big
bubble which shrinks while the others coarsen. The bubbles coarsen
under a target area constraint. These target areas are roughly the
areas each bubble will have at the end of coarsening. We defined
the pinning field above. The Hamiltonian is:
\begin{equation}
H=\sum_{i,j (sites)}\bigg(J - \frac{\gamma}{r_i}\bigg)(1-\delta_{\sigma_i \sigma_j})+
  \sum_{k (bubbles)}\Lambda_k(S_k - S_k^t)^2 \textrm{,}
\end{equation}
where we have generalized the area constraint. The area coupling constant $\Lambda_k$ is:
\begin{equation}
\Lambda_k = \lambda \bigg(\frac{1}{S_k^t}\bigg) S^t _{max} \textrm{,}
\end{equation}
where $\lambda$ is a constant and $S^t _{max}$ is the largest of
the target areas. With a uniform $\Lambda$ as in previous
applications, the area term is larger for bigger bubbles than for
smaller bubbles. Therefore, the former "swallow" the latter in
order to minimize energy. To overcome this problem, we weight the
coupling constant by making it inversely proportional to the
bubble target area.

To estimate the tentative target area, we used the area of a
circle centered at the bubble's calculated center with radius
equal to the distance to the closest vertex. Figure~\ref{maps}b
illustrates this procedure. Clearly this approximation works
better for isotropic bubbles. We could also use the average
distance to the neighboring vertices or other estimates. The
geometrical element that fills two-dimensional space with minimum
surface energy is the hexagon. The area of a hexagon that fits
inside a circle is about 10\% smaller than the area of the circle
itself. Therefore, we use 90\% of the areas of the circles in
fig.~\ref{maps}b as the tentative target areas. The method we use
to obtain the tentative target areas plays only a small role, if
any, in the final reconstruction.

To ensure that all the bubbles grow at comparable rates, we
increase their target areas in steps to their full values. During
the first 50 {\it mcs} we set the target areas to 10\% of their full
values, so after 50 {\it mcs}, all the bubbles have grown to 10\%
of their final sizes. After each 50 {\it mcs}, we increase the
target areas by 10\%, so the target areas reach their full values
after 500 {\it mcs}.

After all the bubbles have achieved their final areas and the
structure has pinned, we turn off the area constraint and impose a
cutoff on the pinning field, as discussed above, to allow the
interfaces to relax, and the areas to change, in order to obtain
the lowest energy configuration.

For many purposes the approximate reconstruction is adequate.
However, if we wish to measure the growth rates of bubbles to high
accuracy we are examining small differences of large areas so even
a small error in the areas gives a large error in their growth
rates. We are developing procedures to correct the bubbles'
areas~\cite{vasconcelos1}.

\subsection{Results}

\begin{figure}[ht]
\centering
\includegraphics[width=\linewidth]{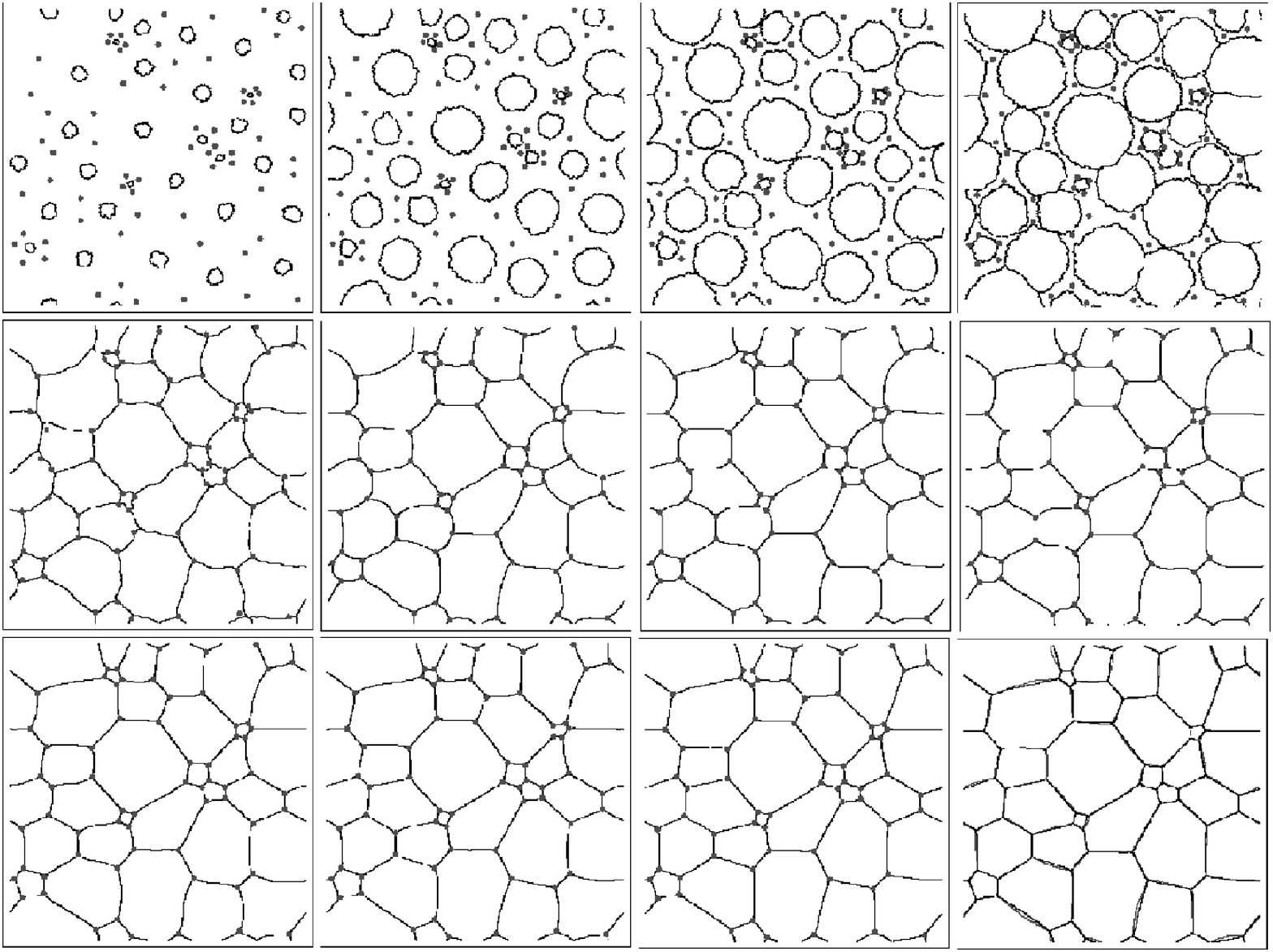}
\caption{Growth of bubbles from their centers for the 50000 {\it mcs} coarsened foam. In the first stages the bubbles have energetically
favorable circular shapes and grow at the same pace. Eventually they impinge on one another forming interfaces and vertices and the vertices
trap on the pinning centers (red dots). The interfaces relax and straighten after we turn off the area constraint and cut off the range of the
pinning field. From the upper left to the lower right corner: 50, 150, 250, 350, 500, 1000, 2000, 3500, 6000, 9000, and 25000 {\it mcs}. The
last picture in the series shows the original (red) and reconstructed (black) structures superimposed. The simulation parameters we
chose are $J=$~1, $T=$~0/1 (see text), $\gamma=$~1000, and $\lambda=$~0.1.}
\label{nuc}
\end{figure}

Figure~\ref{nuc} shows (up to 25000 {\it mcs}) the growth of
bubbles and structures reconstructed for a simulated foam
coarsened for 50000 {\it mcs}. The last picture in the series
shows a snapshot of the simulation superimposed on the
reconstruction. Our reconstructions used the same alternating
temperature approach as our coarsening simulations. We set $T=1$
to allow the vertices to find the pinning centers faster. The
parameters for all reconstruction simulations were $J=$~1, $T=$~1,
$\gamma=$~1000, and $\lambda=$~0.1. We repeated the same procedure
for many different configurations and obtained the reconstructed
foam coarsening of fig.~\ref{recon}.

\subsection{Analysis of Results}

During nucleation and growth of bubbles, the bubbles are at first
circular and grow at the same pace due to the rescaling of their
target areas. Eventually they impinge on one another, forming
interfaces and vertices. The contact between bubbles keeps them
from growing further and both the area constraint and pins further
inhibit global coarsening.

Once the bubbles fill the lattice, their configuration is already
close to the final one. Most of the vertices have already trapped
on pinning centers and most untrapped vertices eventually trap
during the relaxation of edges and rearrangement of areas after we
turn off the target area constraint and impose cutoffs on the
pinning field. The residual pinning field hinders further
coarsening.

Although defects in the center calculation cause most errors in
the reconstructed pattern, the nucleation of bubbles and
relaxation of edges can introduce additional imperfections. A
particular vertex may attach to the wrong pin or an edge may be
missing. A bubble which was originally very long and thin might
receive a target area much different from its original area
because the anisotropy makes the nearest vertex much closer to the
center than the average distance to the vertices. Since all
bubbles nucleate as circles, an edge may miss a pin during the
nucleation of a long, thin bubble, leaving a pin inside the bubble
which never captures the edge. After this bubble achieves its
final target area, it will still have to jump over an energy
barrier in order to squeeze into a long, thin shape. A pair of
stretched bubbles just before a $T1$ process (topological
transition where a four-fold vertex decays into two three-fold
vertices) in the original image will almost never reconstruct
properly because it is so near the much lower energy state after
the $T1$ that the reconstruction jumps over the real configuration
to the lower energy configuration. However, the real foam pattern
would evolve rapidly to this state so the effective error is
small.

\begin{figure}
\centering
\includegraphics[width=\linewidth]{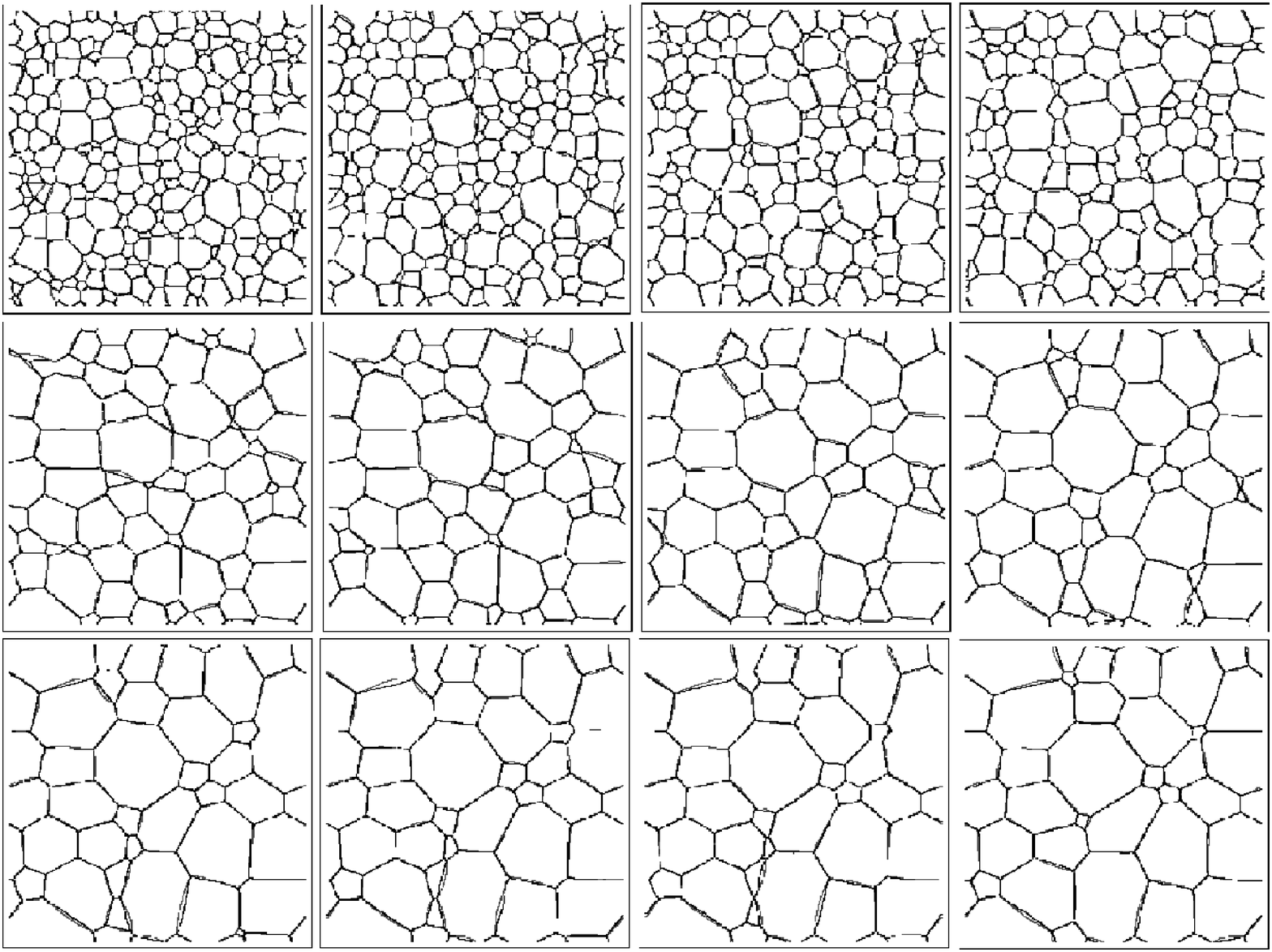}
\caption{Reconstruction (black) of foam evolution during coarsening superimposed on the original structures (red). From upper left to
lower right: 6000, 8000, 10000, 14000, 22000, 26000, 32000, 38000, 42000, 44000, 46000, and 50000 {\it mcs}. For the coarsening procedure
we chose $J=$~1, and $T=$~0/0.5 (see text) while for the reconstruction we chose  $J=$~1, $T=$~0/1, $\gamma=$~1000, and
$\lambda=$~0.1}
\label{recon}
\end{figure}

\begin{figure}[ht]
\includegraphics[width=\linewidth]{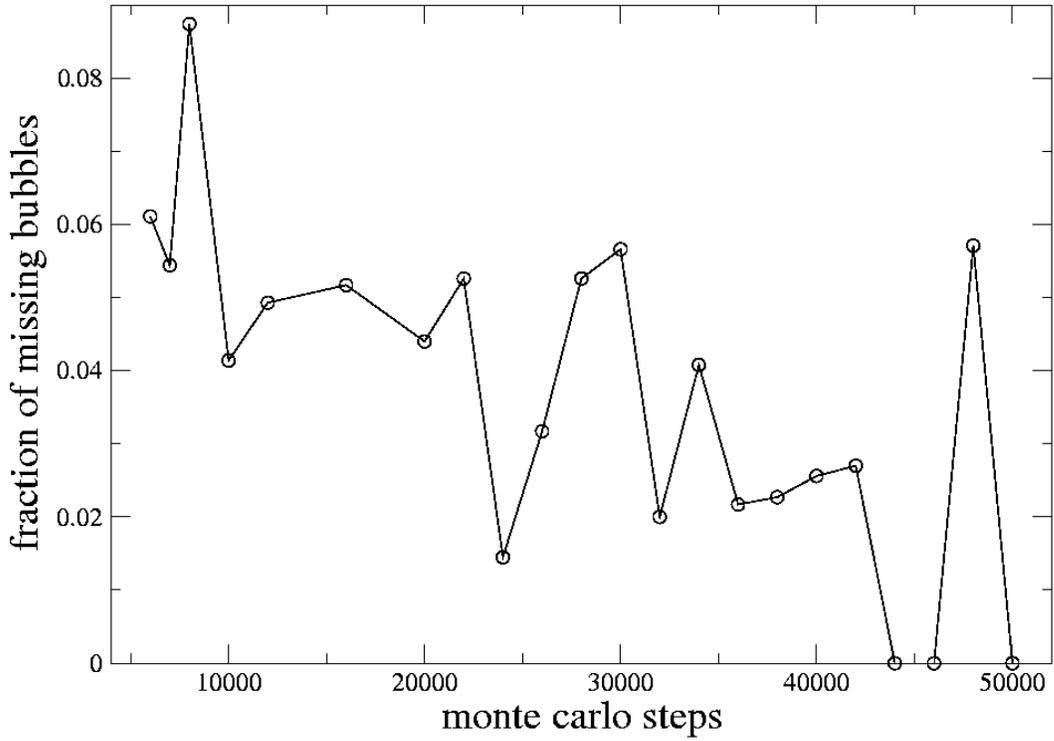}
\caption{Fraction of missing bubbles in the reconstruction as a function of time.}
\label{bub_diff}
\end{figure}

Even after a vertex pins, it may break loose. Vertices in less
energetically favorable configurations, like those undergoing
$T1$s or joining edges that make angles much smaller than
120$^\circ$, may detach from the pins despite very strong coupling
to the pinning field. These extreme situations create loose
vertices that evolve locally with the dynamics of an unpinned foam
which may complete $T1$s in progress or lead to the disappearance of
bubbles. As a result, the reconstructed structure may depart
slightly from the original configuration as in fig.~\ref{recon}.
Errors are less common in more homogeneous, isotropic foams.
Figure~\ref{bub_diff} shows the fraction of missing bubbles as a
function of time.

How do these deviations affect the global coarsening described
in~\cite{srolovitz83,glazier90,glazier87,stavans89}? If we
characterize coarsening by fitting the average area $<A>$ per
bubble as a function of time to a power law:
\begin{equation}
<A> = kt^\alpha \textrm{,}
\end{equation}
we obtain a growth exponent characterizing the coarsening. For a
foam-like material, the characteristic exponent must be
$\alpha=1$~\cite{glazier90}. Besides the mean bubble area, the
basic measures of the state of a foam are the distribution
$\rho(n)$ of the number of sides (the probability that a randomly
selected bubble has $n$ sides), and the normalized area
distribution $\rho(A/<A>)$ (the probability that a bubble has an
area which is a given fraction of the mean bubble area). The
distributions show dynamical scaling, {\it i.e.} they become time
invariant at long times. We define the $m$th moment of the side
distribution as:
\begin{equation}
\mu_m = \sum^{\infty}_{n=2} \rho(n)(n-<n>)^m \textrm{,} \label{moments}
\end{equation}
where $<n>$ is the average number of sides of a bubble in the
pattern (for infinite patterns $<n>=$ 6). The second moment,
$\mu_2$, measures the r.m.s. width of the distribution while
$\mu_3$ and $\mu_4$ measure the asymmetry and the flatness of the
side distribution.

\begin{figure}[ht]
\includegraphics[width=\linewidth]{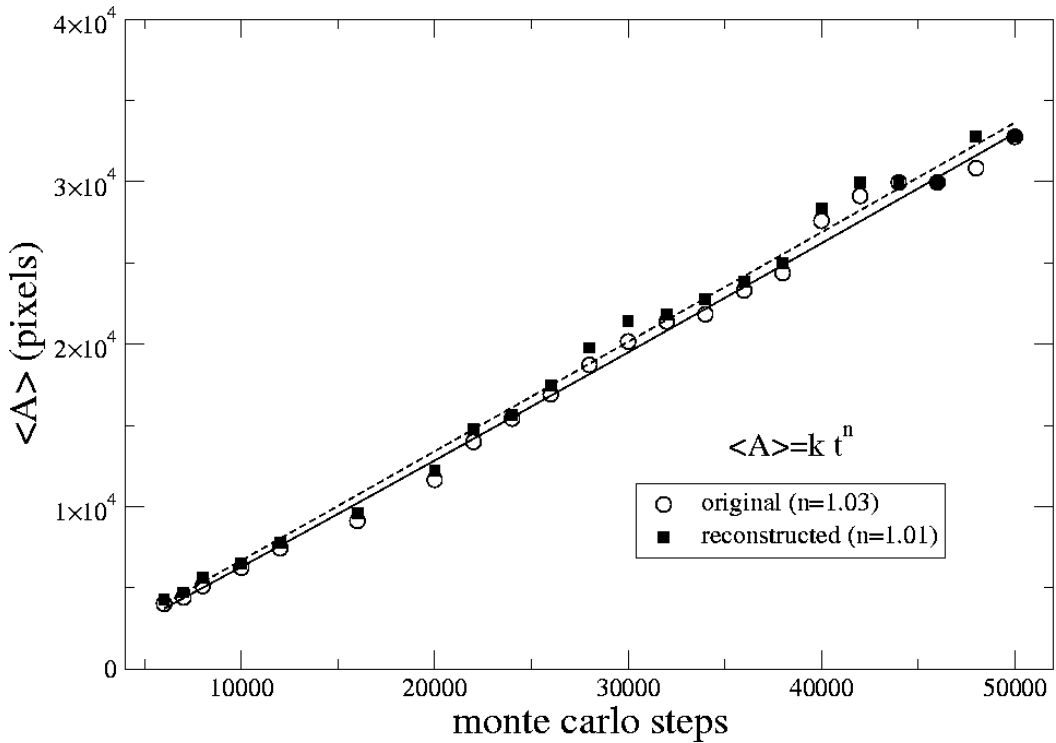}
\caption{Average bubble area evolution for the original and reconstructed structures.}
\label{area_ave}
\end{figure}

To test these properties and quantify the errors, we analyzed the
coarsening dynamics of both the simulated and reconstructed foams.
Figure~\ref{area_ave} shows the time evolution of $<A>$ for both
foams. Both curves follow a power law, with exponents $\alpha =
1.03 \pm 0.03$ for the original foam (solid line in the figure)
and $\alpha = 1.01 \pm 0.03$ for the reconstructed foam (dotted
line). Both exponents are consistent, within error, with the
expected value of one. Thus, the reconstructed foam has the same
global dynamics as the original and theory.

\begin{figure}[ht]
\includegraphics[width=\linewidth]{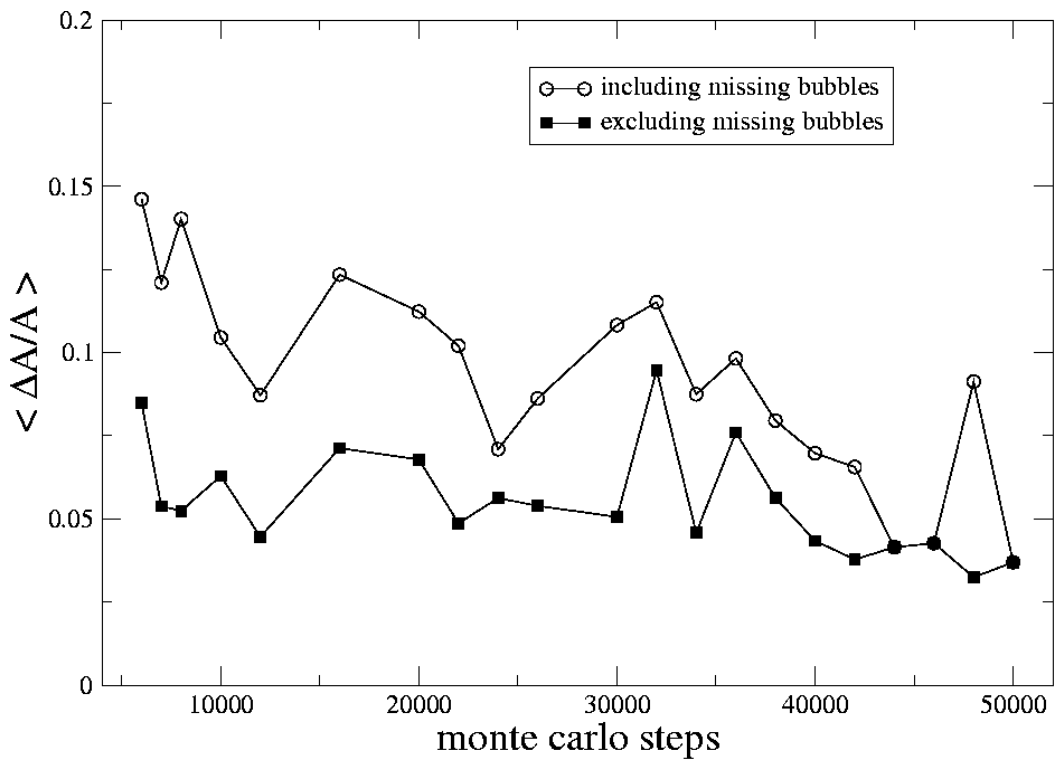}
\caption{Relative mean area deviation of the reconstruction of the simulation as a function of time.}
\label{area_diff}
\end{figure}

Figure~\ref{area_diff} shows the mean deviation of the reconstructed
areas from the original areas given by:
\begin{equation}
\Big<\frac{\Delta A}{A}\Big> = \frac{1}{N}\sum_{i=1}^N \frac{1}{A_i^O}|A_i^O - A_i^R| \textrm{,}
\label{meanarea}
\end{equation}
as a function of time, including and excluding missing bubbles in
the calculation. The superscripts O and R stand for the original
and reconstruction, respectively. The sum in eq.~\ref{meanarea}
runs over all $N$ bubbles in the foam. Both curves decrease
slightly due to the increasing length scale of the pattern. The
time averaged deviation is $(9.2 \pm 3.0) \%$ including missing
bubbles and $(5.4 \pm 1.6) \%$ excluding them. Hence, the area of
a reconstructed bubble differs by about 5.4\% from the area of the
original bubble.

\begin{figure}[ht]
\centering
\includegraphics[width=\linewidth]{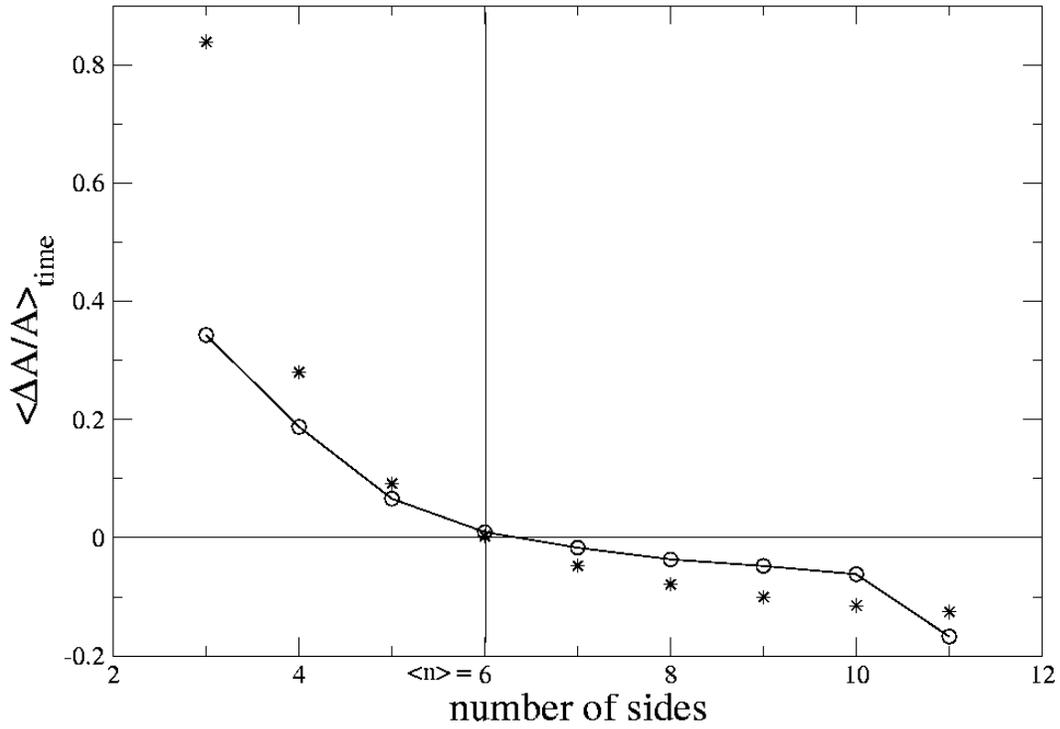}
\caption{Relative difference between simulated and reconstructed areas averaged over time as function of bubble topology (number of sides).
The stars represent the relative deviation of a regular Plateau bubble from a bubble with straight sides  $\delta(P/\sqrt{A})_n$ (see text).}
\label{areatopol}
\end{figure}

\begin{figure}[ht]
\centering
\includegraphics[width=\linewidth]{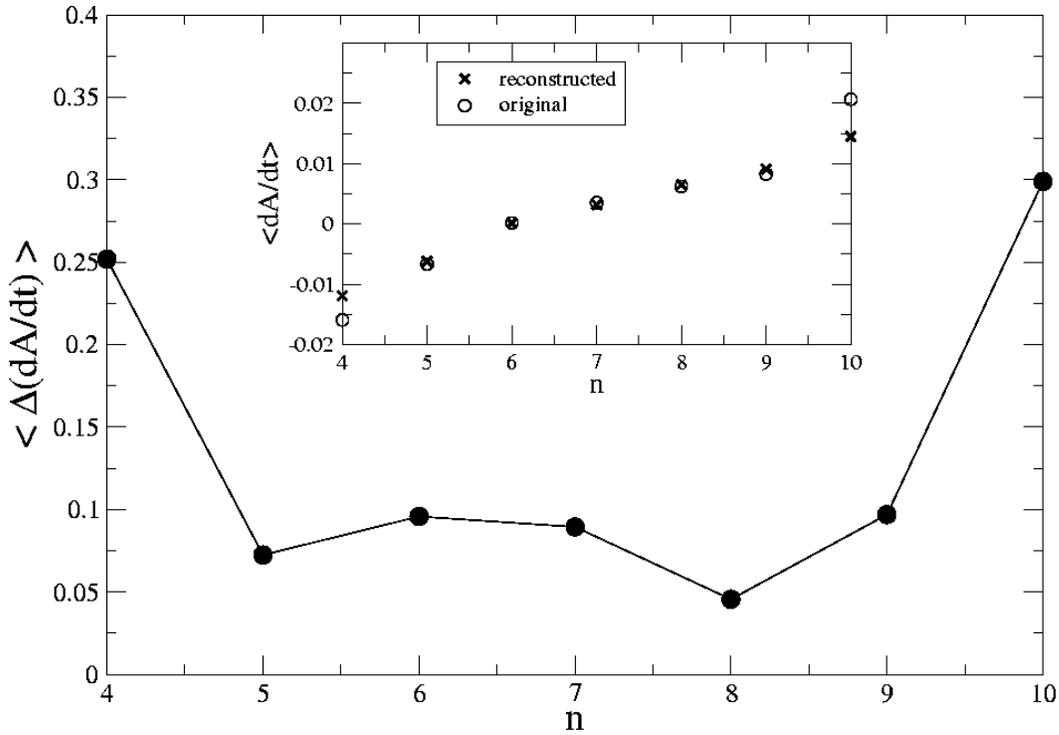}
\caption{von Neumann's law. Detail: average $dA_n/dt$ as a function of $n$ for the simulated and reconstructed foams. Main graph:
percentage difference in von Neumann's law between the simulated and reconstructed foams for each value of $n$.}
\label{figneumann}
\end{figure}

This residual error is a consequence of reconstructing straight
edges rather than the curved boundaries of a real foam. According
to Plateau's rules, the ratio between the perimeter and the square
root of the area of a regular bubble is a constant ($P/\sqrt{A}
\approx 3.72$) while the ratio for polygonal bubbles with straight
sides is $2\sqrt{n\tan(\pi/n)}$. Therefore, the relative deviation
from a regular Plateau bubble of a bubble with straight sides is
approximately $\delta(P/\sqrt{A})_n=2\sqrt{n\tan(\pi/n)} - 3.72$.
Figure~\ref{areatopol} shows $\delta(P/\sqrt{A})_n$ (stars) on top
of the time averaged relative difference between simulated and
reconstructed areas given by:
\begin{equation}
\Big<\frac{\Delta A}{A}\Big>_{time} = \frac{1}{N_n}\sum_{i=1}^{N_n} \frac{1}{A_i^O}(A_i^O - A_i^R) \textrm{,}
\label{eqareatopol}
\end{equation}
as a function of the number of sides $n$. The sum in
eq.~\ref{eqareatopol} runs over all $N_n$ bubbles with $n$ sides
at any instant of time. For few-sided bubbles, the reconstructed
area is smaller than the original area, while for many-sided
bubbles the reconstructed area is larger than the original area.
The reconstructed images had no 3- or 11-sided bubbles to analyze.
Otherwise, the relative area deviations follow very closely the
theoretical predictions for $\delta(P/\sqrt{A})_n$.

von Neumann's law states that, in two dimensions, the area growth rate of
individual bubbles with $n$ edges is proportional to $(n-6)$:
\begin{equation}
\frac{dA_n}{dt} = \kappa (n-6) \textrm{.}
\label{vonneumann}
\end{equation}
Few-sided bubbles are convex, so a straight-sided reconstruction
cuts off the bulges, while many-sided bubbles are concave so a
straight-sided reconstruction fills in the dips.
Figure~\ref{figneumann} shows the deviation in von Neumann's law
calculated for the simulated and reconstructed foams as:
\begin{equation}
\Big<\Delta \Big(\frac{dA_n}{dt}\Big)\Big>= \frac{1}{N_n}\sum_{i=1}^{N_n} \Big(\frac{dA_i^O}{dt}\Big)^{-1} \Big|\frac{dA_i^O}{dt}
        -\frac{dA_i^R}{dt}\Big| \textrm{,}
\label{eqneumann}
\end{equation}
where the summation in $i$ runs over all the $N_n$ $n$-sided
bubbles at any instant of time. The nonlinearity seen in the inset
to fig. ~\ref{figneumann} results from deviations in growth rate
for small bubbles which are an artifact of the regular Potts Model
simulation~\cite{holm91}.

\begin{figure}[ht]
\includegraphics[width=\linewidth]{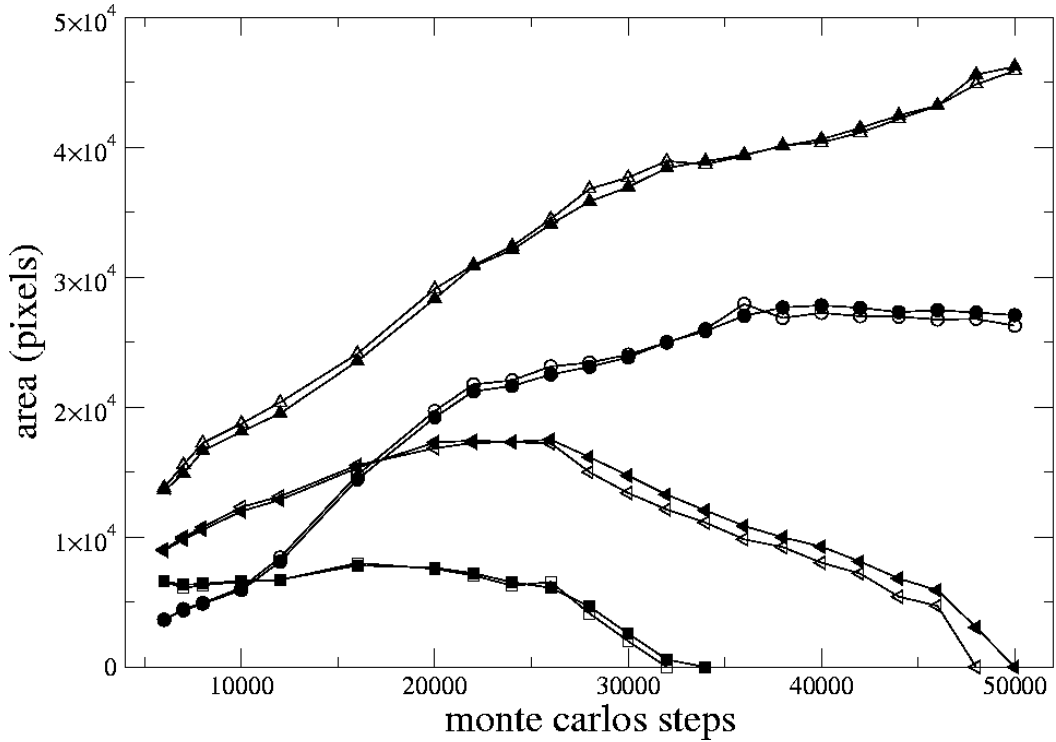}
\caption{Area evolution for four simulated (filled symbols) and reconstructed
(hollow symbols) bubbles.}
\label{area_bub}
\end{figure}

Figure~\ref{area_bub} shows the time dependence of the areas of
four individual bubbles. Filled and open symbols of the same type
(circles, squares, {\it etc.}) correspond to simulated and
reconstructed areas. The local dynamics are very similar for the
simulation and reconstruction. Figure~\ref{recon} shows that for
the great majority of the bubbles, simulation and reconstruction
follow the same dynamics.

\begin{figure}[ht]
\centering
\subfigure[original]       {\includegraphics[width=0.45\linewidth]{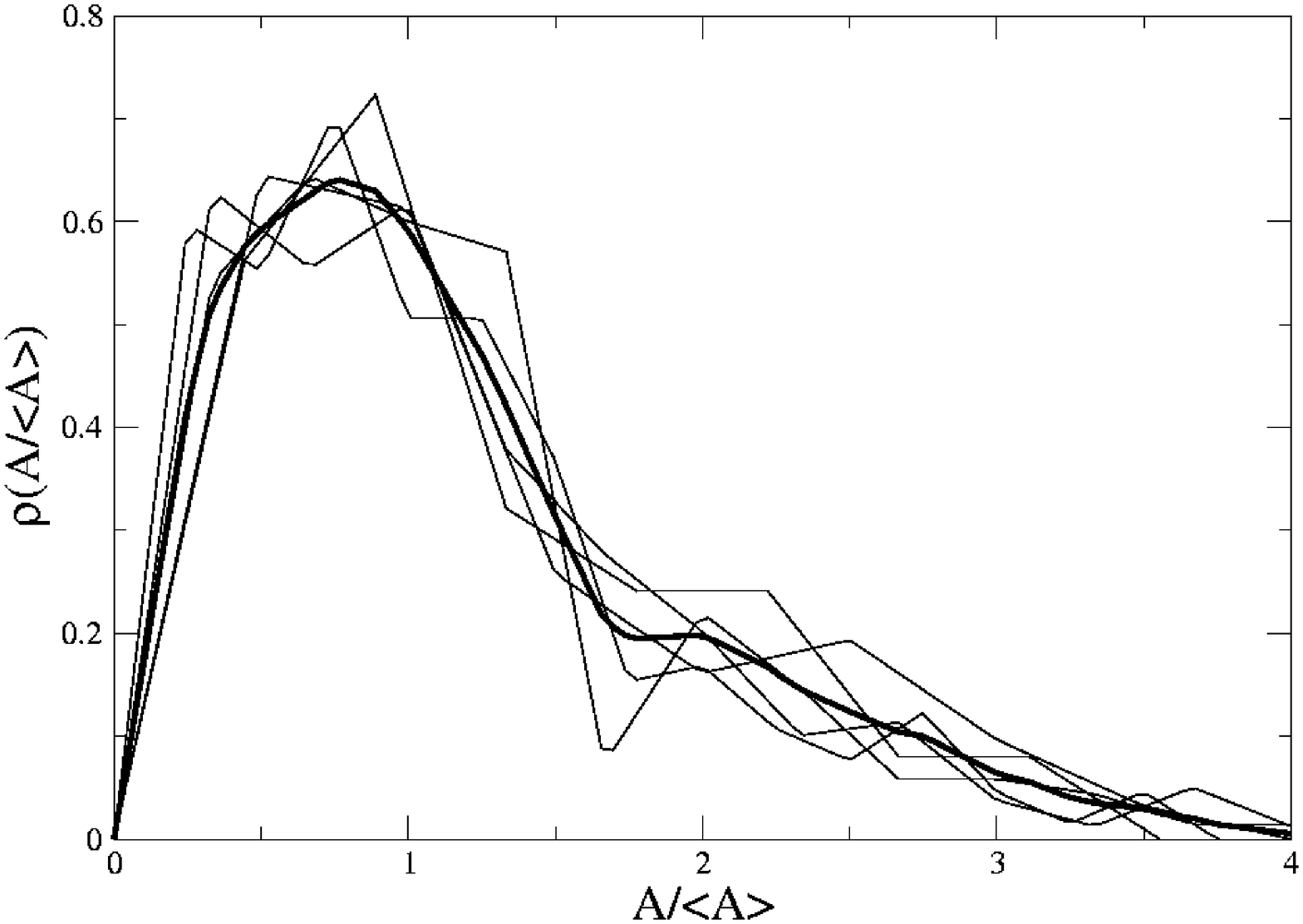} }
\subfigure[reconstruction] {\includegraphics[width=0.45\linewidth]{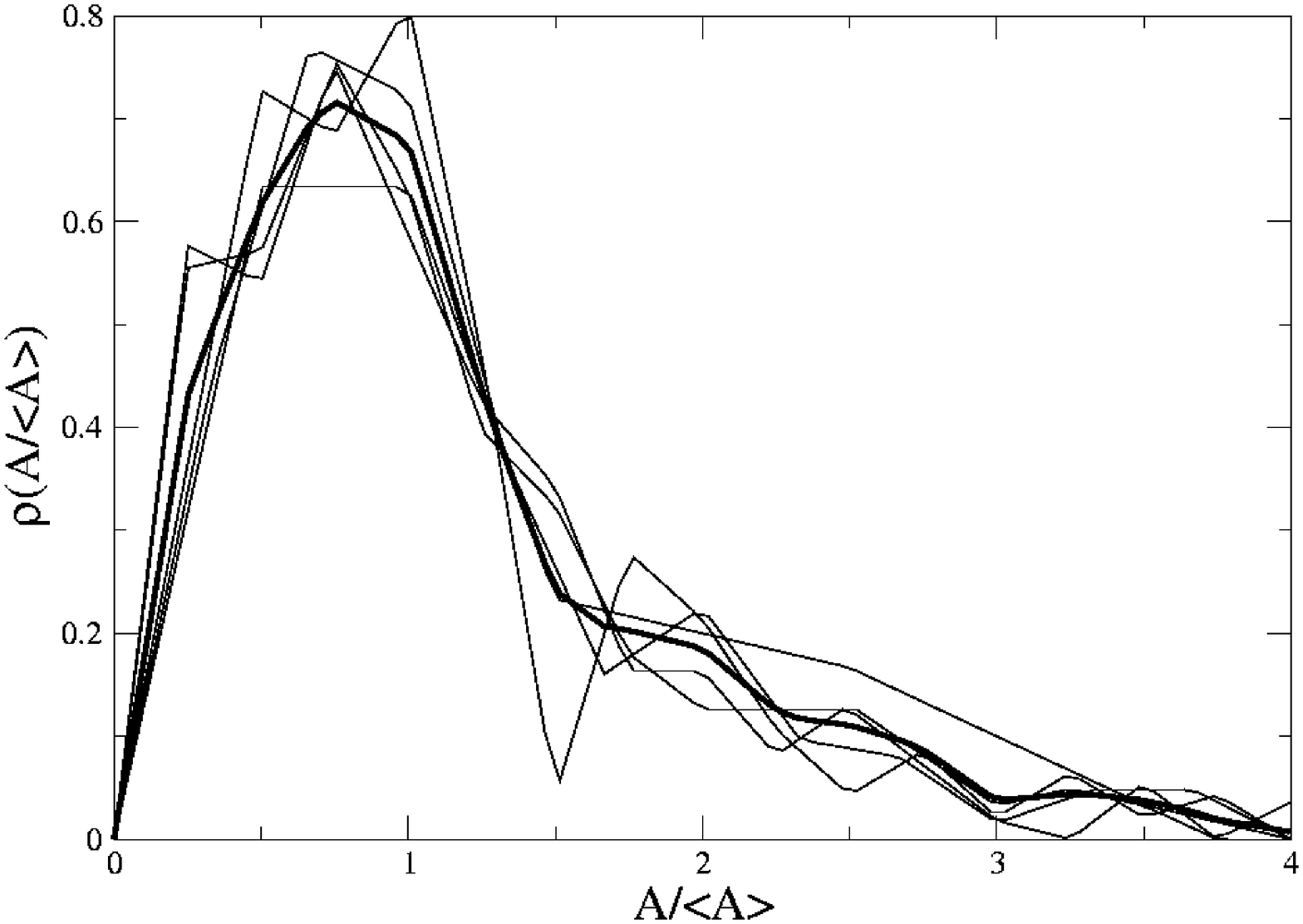} }
\caption{Normalized area distributions for the simulated (a) and reconstructed (b) foams. Each graph shows several distributions for
different times. All distributions have approximately the same shape (thick line).}
\label{dist}
\end{figure}

\begin{figure}[ht]
\centering
\includegraphics[width=\linewidth]{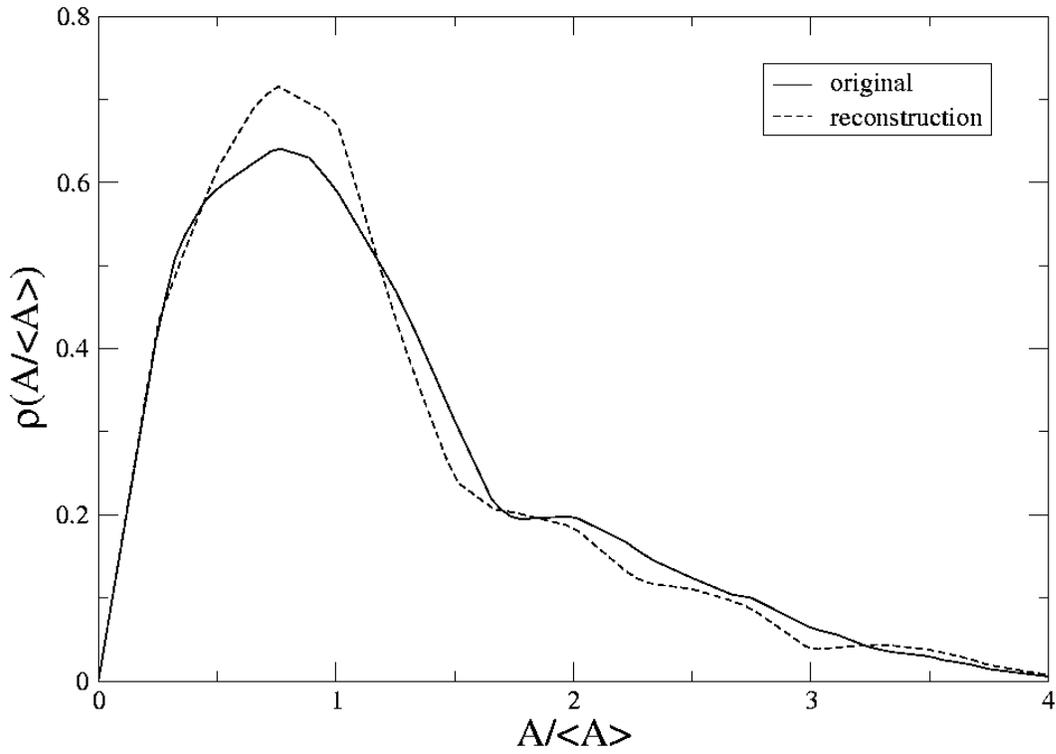}
\caption{Comparison of the shape of the simulated and reconstructed area distributions.}
\label{distcomp}
\end{figure}

Another important aspect of coarsening foam is self-similarity.
Figure~\ref{dist}a shows the normalized area distributions at
several times for the simulated foam. These distributions are
congruent, indicating that the foam is in a scaling state. The
thick line obtained by averaging all the distributions represents
all the individual cases within $3\%$. The same applies to the
reconstruction (fig.~\ref{dist}b). The scaled area distributions
are equivalent within $7\%$ (fig.~\ref{distcomp}) so the
reconstructed foam scales in the same way as the simulation.

\begin{figure}[ht]
\centering
\subfigure[original]       {\includegraphics[width=0.45\linewidth]{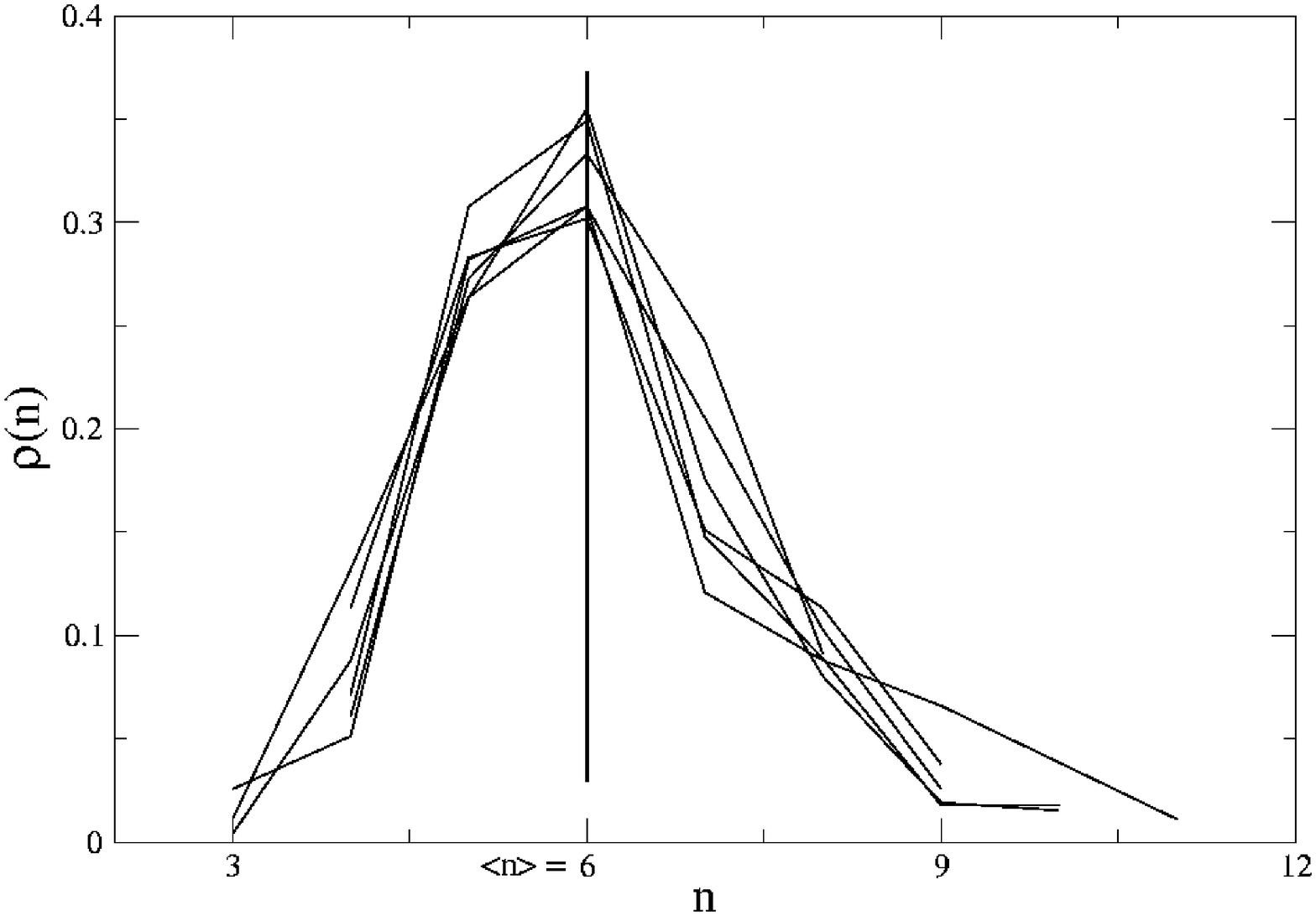} }
\subfigure[reconstruction] {\includegraphics[width=0.45\linewidth]{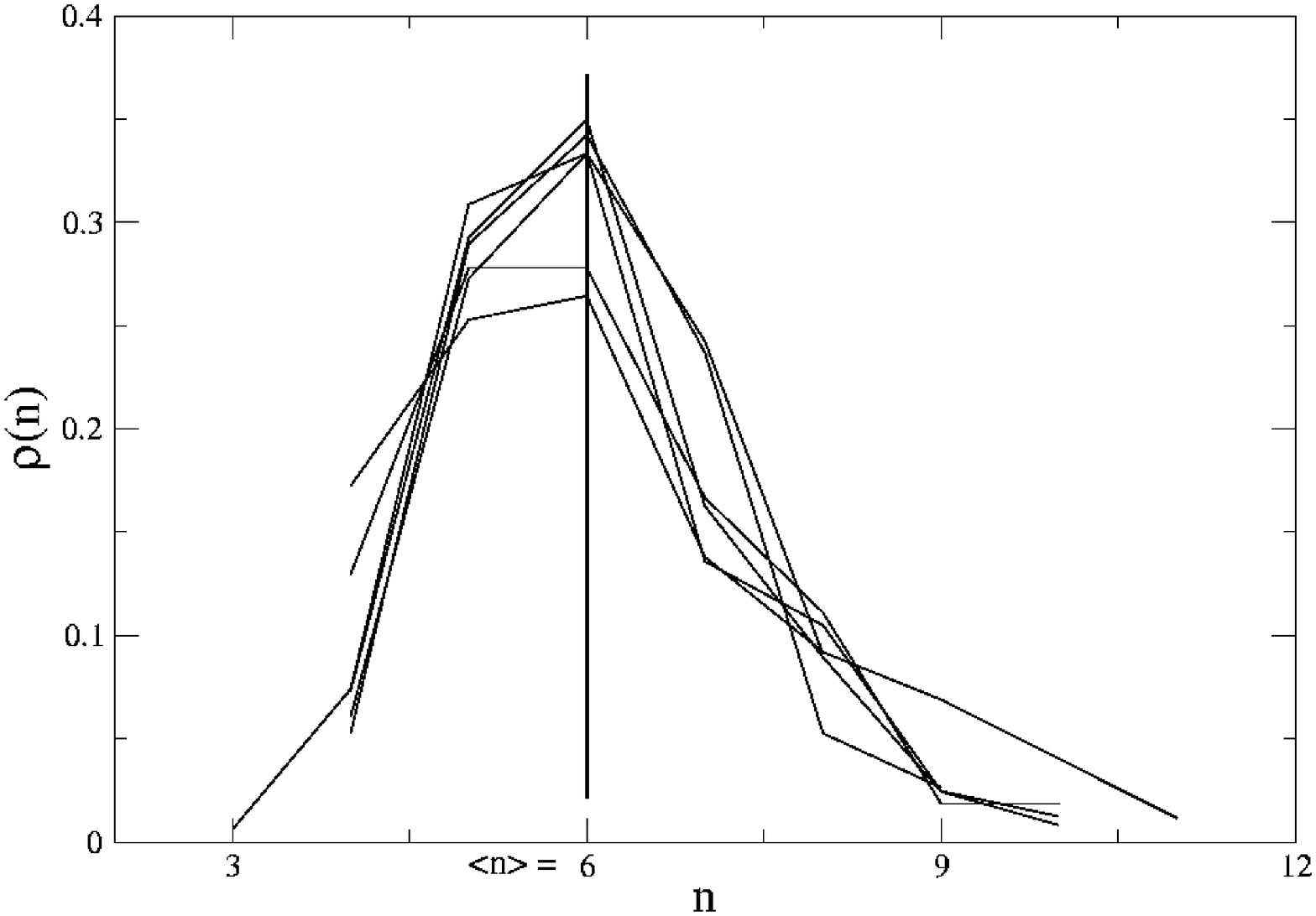} }
\caption{Side distribution $\rho(n)$ as a function of time for the simulation
(a) and reconstruction (b).}
\label{facedist}
\end{figure}

\begin{figure}[ht]
\includegraphics[width=\linewidth]{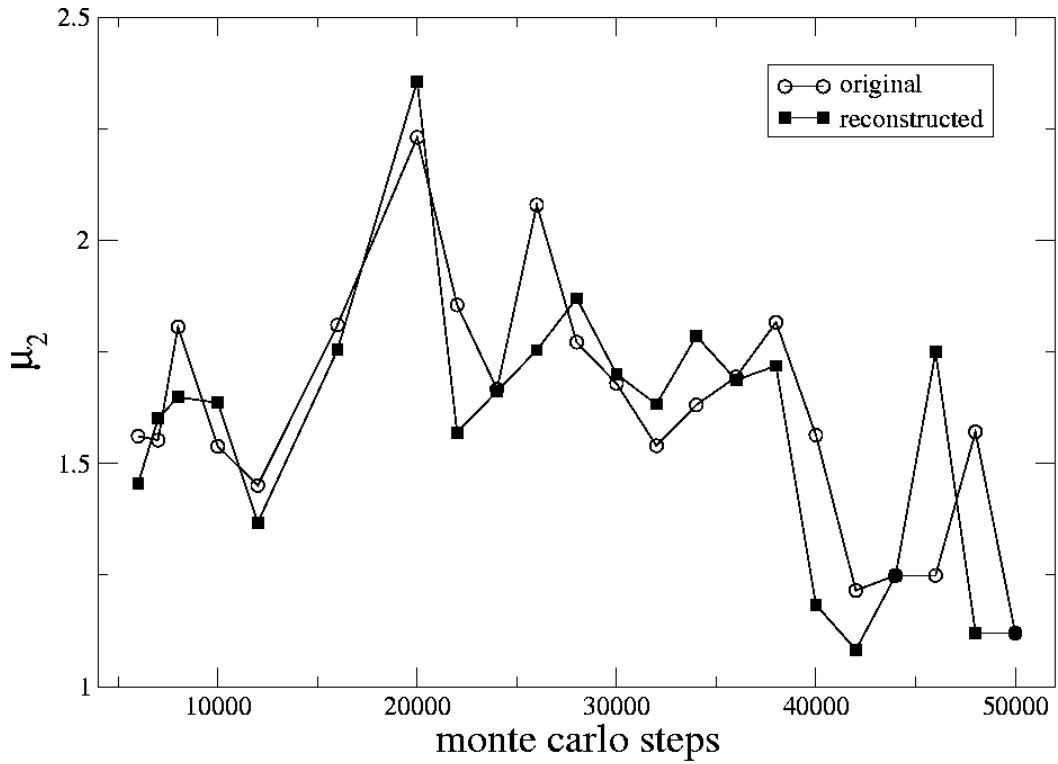}
\caption{Second moment $\mu_2$ of the side distribution as a function of time
for the simulation  and reconstruction.}
\label{figmu2}
\end{figure}

Figure.~\ref{facedist} shows the evolution of the side
distributions for both the simulation (a) and reconstruction (b).
The distributions are time-independent and remarkably similar,
peaking at $n=6$ ~\cite{stavans89}. Figure~\ref{figmu2} shows the
second moment of the side distributions as function of time. The
general behavior is similar for both patterns: $\mu_2$ oscillates
about a stable value, since in the scaling state, the width of the
distribution is nearly constant. The average values of $1.62 \pm
0.27$ for the simulation and $1.46 \pm 0.30$ for the
reconstruction agree well with experimental
values~\cite{stavans89,glazier90}.

\begin{figure}
\includegraphics[width=\linewidth]{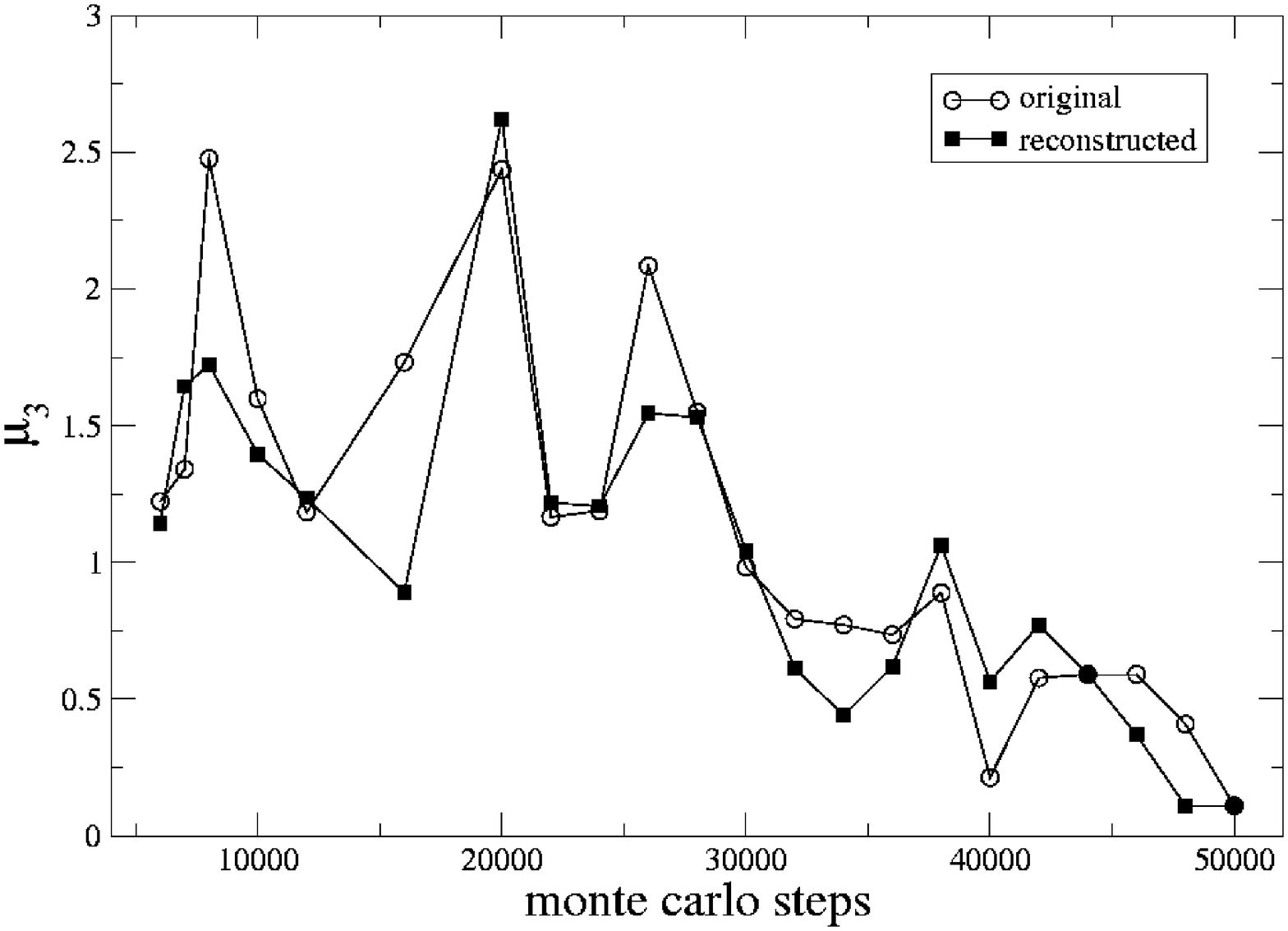}
\caption{Third moment $\mu_3$ of the side distribution as a function of time for
the simulation and reconstruction.}
\label{figmu3}
\end{figure}

\begin{figure}
\includegraphics[width=\linewidth]{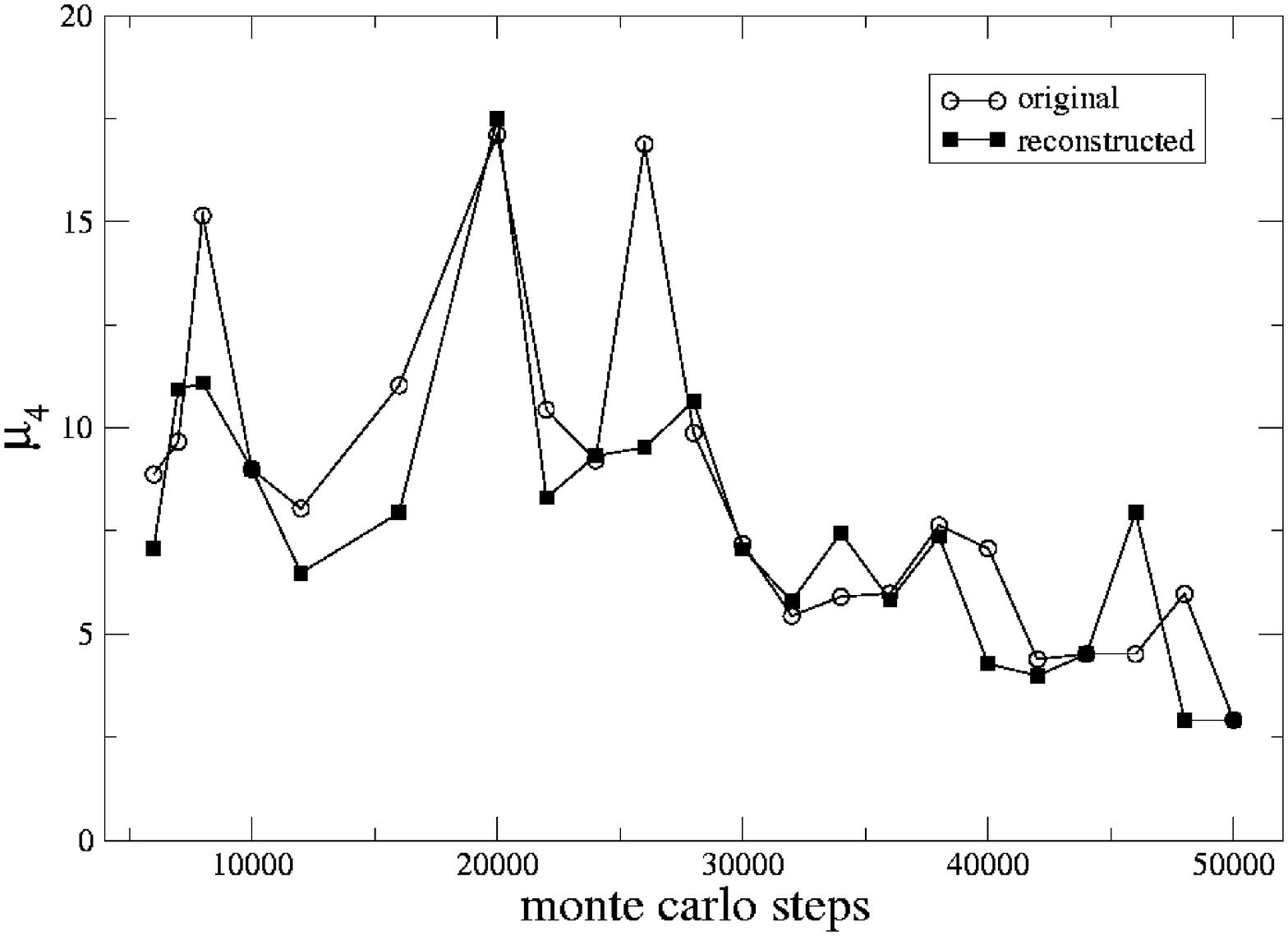}
\caption{Fourth moment $\mu_3$ of the side distribution as a function of time
for the simulation and reconstruction.}
\label{figmu4}
\end{figure}

Figures~\ref{figmu3} and~\ref{figmu4} show, respectively, the
third and fourth moments as functions of time. The third moment
begins near its maximum ($\mu_3= 2.5$) showing the inhomogeneity
in the foam that lasts until about 30000 {\it mcs}. It then drops
to a stable value of $\mu_3= 0.61 \pm 0.27$ for the simulation and
$\mu_3= 0.63\pm 0.33$ for the reconstruction. The fourth moment
behaves similarly: an initial equilibration stage up to 30000 {\it
mcs}, followed by stability with $\mu_4= 5.6\pm 1.4$ for the
simulation and $\mu_4= 5.5 \pm 1.8$ for the reconstruction, again
agreeing with experiment~\cite{glazier90}.

\section{Conclusions}

Our method of reconstructing two-dimensional foam structures works
well. The Potts Model generates minimal energy surfaces and the
pins hinder coarsening. These two tools together reconstruct a
coarsening foam. We show that imperfections mainly result from the
distance map calculation, which is not the core of our method. A recent paper of Kammerer and
Glantz suggests an alternative scheme which may eliminate many of
these errors~\cite{kammerer03}. The
deviations do not induce systematic errors in the statistical
properties of a foam during coarsening. The method extends to
three dimensions with no major modifications, which will be very
helpful in studies of the dynamics of three-dimensional cellular
patterns.

Our aims are now two-fold. We must deal with real data. The
reconstruction algorithm presented in this article will be the
basis for analyzing the images of three-dimensional foams we
recently obtained using X-ray tomography at the European
Synchrotron Radiation Facility (ESRF). At this point, we stress
that the major difference between two-dimensional and three-dimensional partial images is that
the former consists of vertices while the later consists of
edges of bubbles.

To extract the maximum information from experimental images we
will directly couple the value of $J$ in the Potts Hamiltonian to
the pixel gray level. A bubble boundary located in a dark region
(fluid phase) will cost much less energy than a boundary in a
light region (gas phase). This reformulation of the pinning field
has the great advantage that it requires no image pretreatment or
thresholding. Randomly distributed ``fluid'' pixels due to
experimental noise will have much less power to pin boundaries
than the organized pixels corresponding to edges, even if their
gray level values are of the same order of magnitude.
The method is also relatively immune to long wavelength
inhomogeneities in the image saturation, {\it e.g.} due to MRI
coil geometry. 

We are working to improve our image
reconstruction by developing heuristics to correct for
missing/extra bubbles, by introducing additional mechanisms to
deal with noise, and by implementing target area corrections.

\ack We thank Dr. Mark Zajac and Dr. Marius Asipauskas for great
help and infinite patience in our countless discussions, Dr. Igor
Veretennikov and Dr. Burkhard Prause for the MRI image of a foam
and Prof. Fran\c cois Graner for his helpful revisions to the
manuscript. We acknowledge support from the Center for Applied
Mathematics of the University of Notre Dame and NSF grants
DMR-0089162 and INT98-02417, NASA grant NAG3-2366 and DOE grant
DE-FGO299ER45785.


\end{document}